\documentclass[12pt]{iopart}
\usepackage{graphicx}
\usepackage{dcolumn}
\usepackage{bm}

\usepackage[numbers]{natbib}
\usepackage{placeins}

\usepackage[english]{babel}

\usepackage{epstopdf}

\usepackage{hyperref}

\usepackage{color}

\newcommand{\drm}{\mathrm{d}}
\newcommand{\Etot}{E_{\rm{tot}}}

\newcommand{\xn}{x_{n  }}
\newcommand{\xnmo}{x_{n-1}}
\newcommand{\xnmt}{x_{n-2}}
\newcommand{\xnpo}{x_{n+1}}

\begin{document}
\title[Power law error growth in multi-hierarchical chaotic systems]{Power law error growth in multi-hierarchical chaotic systems -- a
  dynamical mechanism for finite prediction horizon in weather forecasts}
\author{Jonathan Brisch and Holger Kantz}
\address{Max Planck Institute for the Physics of Complex Systems (MPIPKS), 01187 Dresden, Germany}
\ead{brisch@pks.mpg.de, kantz@pks.mpg.de}

\begin{abstract}
We propose a dynamical mechanism for a scale dependent error growth
rate, by the introduction of a class of hierarchical models. The coupling of
time scales and length scales is motivated by atmospheric dynamics. 
This model class can be tuned to exhibit a scale dependent error growth rate in
the form of a power law, which translates in power law error growth over time 
instead of exponential error growth as in conventional chaotic systems. 
The consequence is a strictly finite prediction horizon, since in the limit of
infinitesimal errors of initial conditions, the error growth rate diverges 
and hence additional accuracy is not translated into longer
prediction times. By re-analyzing data of the NCEP Global Forecast System
 published by Harlim
et al.\cite{Harlim2005} we show that such a power law error growth rate can
indeed be found in numerical weather forecast models.
\end{abstract}

\noindent{\it Keywords\/}:chaos, finite prediction horizon, atmospheric dynamics

\submitto{\NJP}

\maketitle

\section{Introduction}
There is a long (but sparse) debate in the community of atmospheric physics
and meteorology about the prediction horizon of weather forecasts. As it was
prominently pointed out by Lorenz (1963)\cite{Lorenz1963}, the
chaotic nature of the atmosphere when seen as a dynamical system 
has the consequence of sensitive
dependence on initial conditions. Hence, even infinitesimal errors in the
initial condition of a forecast model compared to the real atmospheric state
grow exponentially fast in time and eventually reach macroscopic
scales. Then the model forecast has no similarity to the real state
anymore, and the forecast time after which this is the case is called the
prediction horizon. As argued in \cite{Bauer2015}, this horizon has been pushed
forward by about 1 day/decade in the past 3-4 decades, and is now, with current
observation technology including remote sensing, current data assimilation, 
nowadays physical understanding of the atmospheric processes, 
and computer power, at around 10
days. It is also argued in \cite{Bauer2015} that this progress will continue 
and that one day one might be able to perform 
{\sl multi-seasonal} weather forecasts
with high-resolution models.

Indeed, this assumption is compatible with 
the conventional idea of {\sl exponential
  error growth} defined by positive Lyapunov exponents\cite{ott_book}: 
An initial perturbation
of a state vector of size $E_0$ grows in time $t$ as
$E(t)=E_0 e^{\lambda t}$, where $\lambda >0$ is the largest
Lyapunov exponent of the system. A reduction of $E_0$ by $1/e$   
(by more accurate/ more complete observations of the current state) will
extend the prediction horizon linearly by one Lyapunov time $1/\lambda$,
\begin{equation}
	E(t) = E_0 \rm{e} ^{\lambda t},\;\;\;  t_{\rm pred} =
        \frac{1}{\lambda}\left( \ln(E_\infty) - \ln(E_0)\right) \rightarrow
        \infty\;\mbox{ for }{E_0\to 0},
\end{equation}
where $E_\infty$ is the diameter of the attractor, which is the saturation
amplitude of any errors and which means complete loss of information about the
true trajectory at time $t_{\rm pred}$, the prediction horizon. 
Even if it is commonly assumed that reducing initial errors by orders of
magnitude for a linear gain in prediction horizon is infeasible, and hence
this classical notion of chaos usually implies unpredictability in the long, 
at least in principle there is no limit to the prediction horizon.

Since long there have been warnings in the atmospheric physics literature that
 error growth might be dramatically different here, starting from
Thompson\cite{THOMPSON1957} in 1957, Robinson\cite{Robinson1967} in 1967, and 
Lorenz\cite{Lorenz1969} in 1969. In a recent paper Palmer et
al. \cite{Palmer2014} coined the notion of the 'the \textit{real} butterfly
effect' for a strictly finite prediction horizon.
In Refs. \cite{THOMPSON1957,Lorenz1969,Robinson1967,Robinson1971,Leith1972}
the authors investigated the Navier-Stokes equation or some similar empirical
flow equation in two and three-dimensions, with and without dissipation and
different energy spectra ranging from $ E(k) \sim k^{-5/3}  $ to $ k^{-3}
$. They all conclude that there exists a fundamental limit of predictability
which is an \textit{intrinsic} property of the investigated flow
equation. Applying  the results to the atmosphere by setting similar energy
spectra and time and length scales the authors conclude that this fundamental
limit of predictability of the atmosphere lies between 7 days
\cite{THOMPSON1957}, 10 days \cite{Robinson1967} and approx 14 days
\cite{Lorenz1969,Smagorinsky1969}.  The recent study of ECMWF \cite{Zhang2019}
on weather forecast systems comes to an intrinsic limitation of 15 days.

\begin{figure}
	\centering
	\includegraphics[width=\columnwidth]{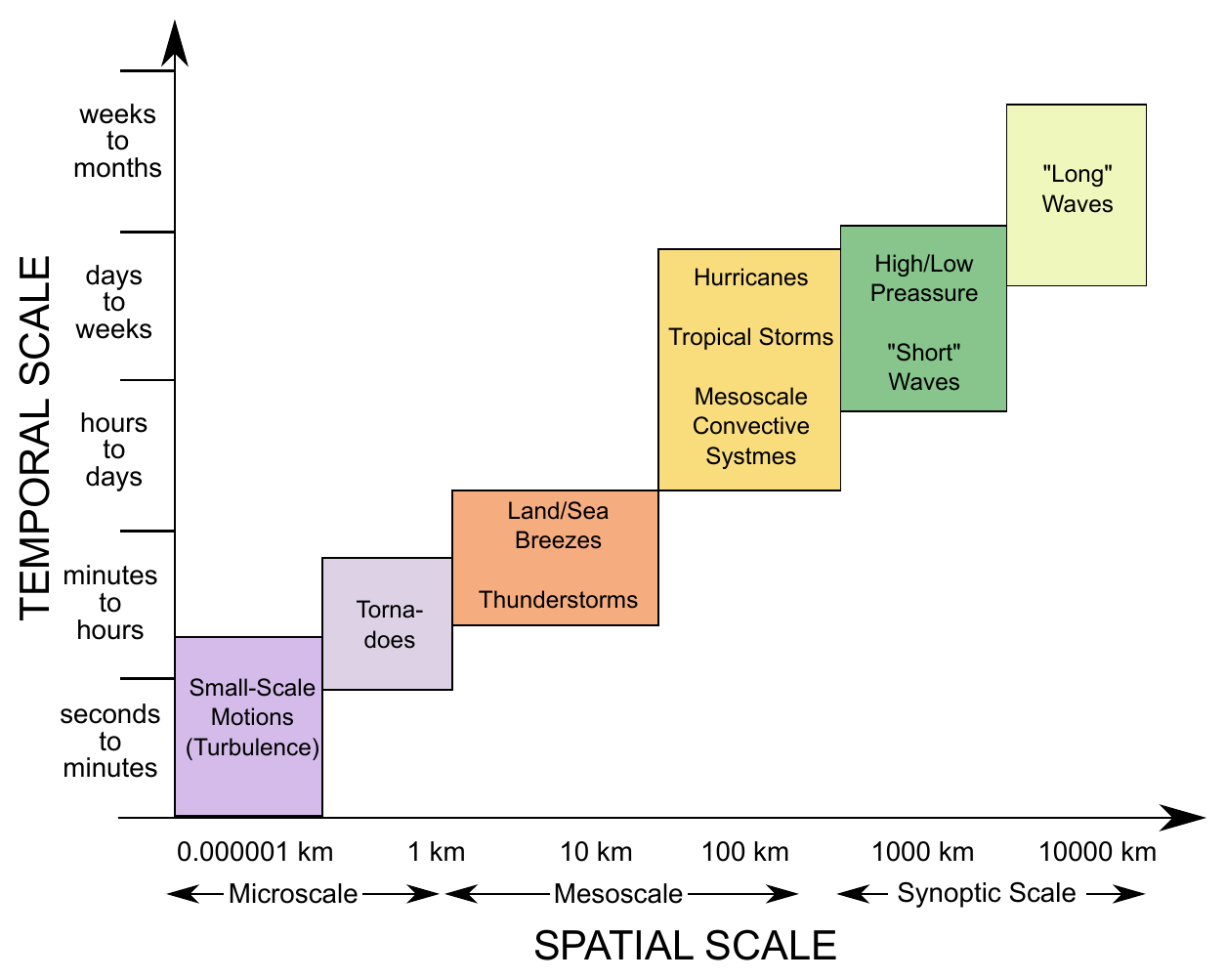}
	\caption{Typical meteorological graph about time and length scales in
          the atmosphere. Own reproduction.} 
	\label{fig:Meteo_plot_atmosphere_time_spatial_scales}
\end{figure}

Atmospheric dynamics takes place on a hierarchy of spatial and
temporal scales which are coupled, see
Figure\ref{fig:Meteo_plot_atmosphere_time_spatial_scales}. Whereas synoptic  
scale structures of
sizes of several 1000km (e.g., high and low pressure systems) live on time
scales of several days, small scale structures such as clouds show dynamics on
the scale of minutes to hours. It is plausible that along with these
life-times, also error growth takes place on different time scales: the
smaller 
the spatial extent of some structure, the faster it evolves, and hence
its prediction might fail correspondingly earlier.
 This has given rise to
the notion of {\sl scale dependent error growth}: The conventional Lyapunov
exponent should be replaced by a scale dependent quantity, e.g., 
a finite size Lyapunov exponent \cite{Cencini2013}. Indeed, in a study of scale
dependent error growth in the Global Forecast System of the National Center
for Environmental Prediction, Harlim et al.~\cite{Harlim2005} 
have shown that there is a scale dependent error
growth rate which becomes very large if the errors become small
(see Figure 1 in \cite{Harlim2005}).

We propose that if the error growth rate with {\sl decreasing} error magnitude
{\sl grows} sufficiently fast, then this will induce a finite prediction
horizon. This behavior could naturally occur in 
systems which are described by partial differential equations PDEs (such as
the Navier Stokes equations), and would require that the dynamics creates a
hierarchy of spatial and temporal scales as it exists in the atmosphere. 
In the mathematical sense, this would imply a maximal Lyapunov
exponent of $\lambda=\infty$, which, however, would be inaccessible in
standard numerical simulations because of coarse graining of the continuum and
thereby cut-offs in the spatial scale. 

In the remainder of this Letter, we will first introduce the idea of a power
law dependence of the error growth rate on the error magnitude and show that
this leads to a strictly finite prediction horizon. We then introduce a 
class of dynamical systems which shows exactly this behavior, and present as
a specific example a hierarchy of coupled Lorenz96-1 models where numerical
simulations validate a power law divergence of the error growth rate. 
Finally, we re-interprete data from the study by
Harlim et al.\cite{Harlim2005} and show that what they observed is a power law
divergence of error growth rates, which becomes evident in our new
presentation of their data. 

\begin{figure}
\centering
\includegraphics[width=\columnwidth]{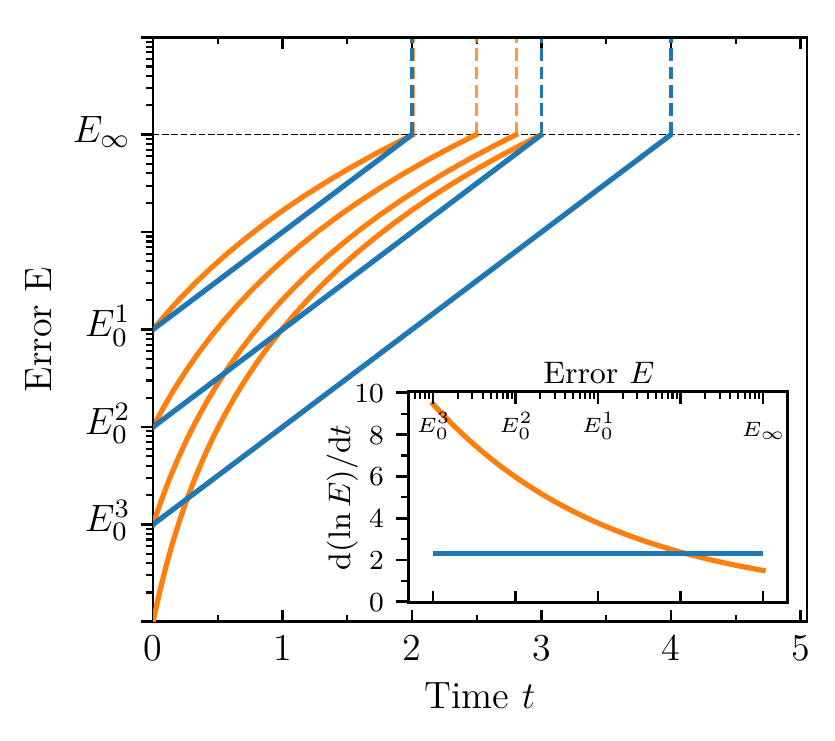}
\caption{Exponential error growth in chaotic systems in blue, power law error
  growth according to (3), prediction horizons indicated by vertical dashed
  lines. The inset shows the divergence of scale dependent
  growth rate $\lambda(E)$ as opposed to a standard Lyapunov exponent.}
\label{fig:Introduction_error_type_gesamt}
\end{figure}

\section{Power law divergence of scale dependent error growth}
Let us assume that the dynamics exhibits a scale dependent error growth rate
$\lambda(E) := \frac{\drm
\ln(E(t))}{\drm t}|_{E(t)=E}$ where the rate of growth is a
power law with an exponent $-\beta$ and some coefficient $a>0$ and where
$E$ is the magnitude of a perturbation:
\begin{equation}\label{eq:2}
	\frac{\drm \ln(E)}{\drm t}  = \frac{\dot{E}}{E} = a E^{-\beta}.
\end{equation}
Integration by separation of variables leads to a power law growth of errors.
\begin{equation}\label{eq.:power-law-error-growth}
	E(t) = (E_0^\beta + a \beta t)^{1/\beta}.
\end{equation}
As for classical exponential error growth,
Equation (\ref{eq.:power-law-error-growth}) becomes invalid for very large times $t$
when the error saturates at a value $E_\infty$ related to the finite extent of
the attractor.  What is strikingly different here is 
the diverging error growth rate for small
$E_0$ and small $t$. 
The linear increment $ \Delta t $ we gain in prediction time
every time we cut the error $E_0$ into half becomes smaller and 
smaller (see Figure  
\ref{fig:Introduction_error_type_gesamt}). The overall prediction time
converges to 
a finite value - a maximum prediction horizon $ t_{\rm{max}} $. If the
tolerable maximal error is denoted by $E_{\rm tol}$ then $t_{\rm max}$ is
 given by 
\begin{equation}\label{eq:4}
	t_{\rm pred} = \frac{E_{\rm tol}^{\beta}-E_0^{\beta}}{a \beta} 
	\rightarrow t_{\rm{max}} = \frac{E_{\rm tol}^{\beta}}{a \beta} <
        \infty \mbox{ for }{E_0\to 0}
\end{equation}

This is a new and severe form of chaos, which we propose to exist in systems
with an infinite dimensional phase space. While we assume that under certain
conditions this will be exhibited by PDEs which intrinsically form cascades
such as the Richardson cascade in turbulence, we present here a paradigmatic
model class  
which is based on coupled low-dimensional systems with a hierarchy of scales
imposed by our choice of parameters. 

Given a chaotic $N$-dimensional dynamical system in terms of an ODE,
$\dot{\vec x} = \vec F(\vec x)$, we introduce a hierarchical coupling by 
defining a family of spatial scaling factors $\alpha_i$ decreasing in $i$ 
and of temporal scaling factors $\tau_i$ increasing in $i$. 
Here $i$ denotes the level of the hierarchy, where $i=1$ is the top level and
$i=L$ the lowest. For the $i$-th
level, we replace
$\vec x$ by $\vec x_i/\alpha_i$ and $t$ by $\tau_i t$ and we obtain the
equations  of motion 
\begin{equation}\label{eq:hier}
\dot{\vec{x}}_i = \tau_i [ \alpha_i \vec F (\vec x_i/\alpha_i) + \vec C(\vec
x_{i+1}, \vec x_{i-1})] \;, 
\end{equation} 
for $i=1,\ldots,L$. Here, $\vec C(\vec a,\vec b)$ denotes a {\sl weak}
coupling term  
which in the simplest case might be linear, $\vec C(\vec a,\vec b)=\vec a + \vec
b$ (also weak global coupling is thinkable). 
For a finite
number of levels $L$, the non-existing coupling inputs $\vec x_0$ and $\vec
x_{L+1}$ are set to zero, and
the system then has $N L$ degrees of freedom. Since coupling is weak 
and if the dynamics $\vec F$ generates just one positive Lyapunov exponent
$\lambda$, the 
hierarchical system has $L$ positive Lyapunov exponents which are
approximately $\lambda_i\approx\lambda \tau_i$. We chose the families of $\alpha_i$
and $\tau_i$ being monotonous in such a way that 
 the top level hierarchy is slow but that the spatial extent of its
attractor and the error saturation value $E_\infty$ is large, and that the
lowest level is the fastest and its phase space range is the smallest. 

For infinitesimal errors, the error growth is governed by the maximum Lyapunov
exponent 
$\lambda \tau_L$ of the fastest time scale, but this error growth saturates at
the scale $\alpha_L$ which is small. Then the second largest Lyapunov exponent
of the second lowest level takes over, till also this error growth saturates
at a scale of $\alpha_{L-1}$, and so on. This way we generate a scale
dependent error growth rate, where the properties of this scale dependence are
tunable.

One specific tuning which generates the proposed power-law-divergence of the
error growth rate is to chose both families of scaling factors in a geometric
way, i.e., $\tau_i= c^i$, and $\alpha_i = d^i$. As we will
demonstrate by the help of the specific example below, the resulting 
power $\beta$ of the scaling of $\lambda(E)\propto E^{-\beta}$ is then $\beta=
\ln c /\ln d$.  Clearly, for finite $L$ this divergence is cut-off by a
maximum rate of $\lambda(E)\le \lambda\tau_L$. 

\section{Multi-hierarchical model L96-H}\label{sec:L96_H_novel_model}
We specify now the general model class by choosing the model L96-1 introduced
in Lorenz (1996)\cite{Lorenz1996} for the dynamics $\vec F(\vec x)$. Its
governing equations, using the notation of \cite{Lorenz1996}, read 
\begin{equation}
	\dot{x}_n  = \xnmo \left(\xnpo - \xnmt \right) - \xn + F
\end{equation}
with $ n = 1 \dots N $ , $ x_n $ cyclic permutable with $ x_{n\pm N} = x_n $,
 and $ F $ a
constant driving force. For $ N > 6 $ and $ F >  8 $ all instances
behave chaotically with increasing positive largest Lyapunov exponent for
increasing $ N $ and $ F $.  
Its equations of motion for some inner level $i$ read
\begin{eqnarray}
		\dot{x}_{n, i}  &=& \tau_i \Big[
		\frac{1}{\alpha_i} x_{n-1, i}\left(x_{n + 1, i} - x_{n - 2,
                    i} \right) - x_{n, i} \\
&& + \alpha_i F_i
		+  x_{n, i + 1} 
		+ \frac{\alpha_{i+1}}{\alpha_{i-1}} x_{n, i - 1}
		\Big]\;.
\end{eqnarray}

The system is $ L N $ dimensional and the state space can be divided
into $ L $ subspaces of dimension $ N $ for each level of the hierarchy. 
We denote the state vector by $ \vec{X} = \{\vec{x}_1, \dots, \vec{x}_L \} $ 
with $ \vec{x}_i \in {\rm I\!R}^N $.
The coupling is bidirectional with
\textit{upwards} coupling from lower to higher level $ x_{n, i + 1}  $ and
with \textit{downwards} coupling $ \frac{\alpha_{i+1}}{\alpha_{i-1}} x_{n, i -
  1} $. The pre-factor $ \frac{\alpha_{i+1}}{\alpha_{i-1}} $ is chosen such
that the downwards coupling has the same magnitude as the upwards
coupling. In the lowest level the undefined scale $\alpha_{L+1}$ is 
chosen as continuation of the sequence of $\alpha_i$. The 
term $\alpha_i F_i +  x_{n, i + 1} +
\frac{\alpha_{i+1}}{\alpha_{i-1}} x_{n, i - 1} $ can be considered as time
dependent driving force $ F_i(t) $. It is important to make sure that $ F_i(t)
> \alpha_i \cdot 8 $ for all times to ensure that each level is chaotic. 


In the following, we will present numerical results of this system for 
the parameters $ N = 7$, $F = 15$ for which the single level dynamics (without
coupling) is chaotic with a maximal Lyapunov exponent of $\lambda\approx 2.66$
and an error saturation value $E_\infty\approx 22$.  

We define the \textit{error} $E(t)$ as the ensemble average of the Euclidean 
distance between a reference trajectory and an initially randomly
perturbed error trajectory with perturbation strength $E_0 $. Thereby we
distinguish between the error of the total system $ \vec{X} $ denoted $
\Etot(t) $ and the error $ E_i(t) $ regarding only the subspace of one single
level $ \vec{x}_i $. The scale dependent \textit{error growth rate} is defined
as the time  derivative of the
logarithm of the error $\frac{\drm \ln E}{\drm t} $ as a function of the error
magnitude $E(t)$ at time $t$. 
 Indeed, one can also study the propagation of the error
from level to level by initial perturbations in selected levels only, which
leads to interesting transient behaviors but eventually converges to error
growth as for a global perturbation.

\begin{figure}
	\centering
	\includegraphics[width=\columnwidth]{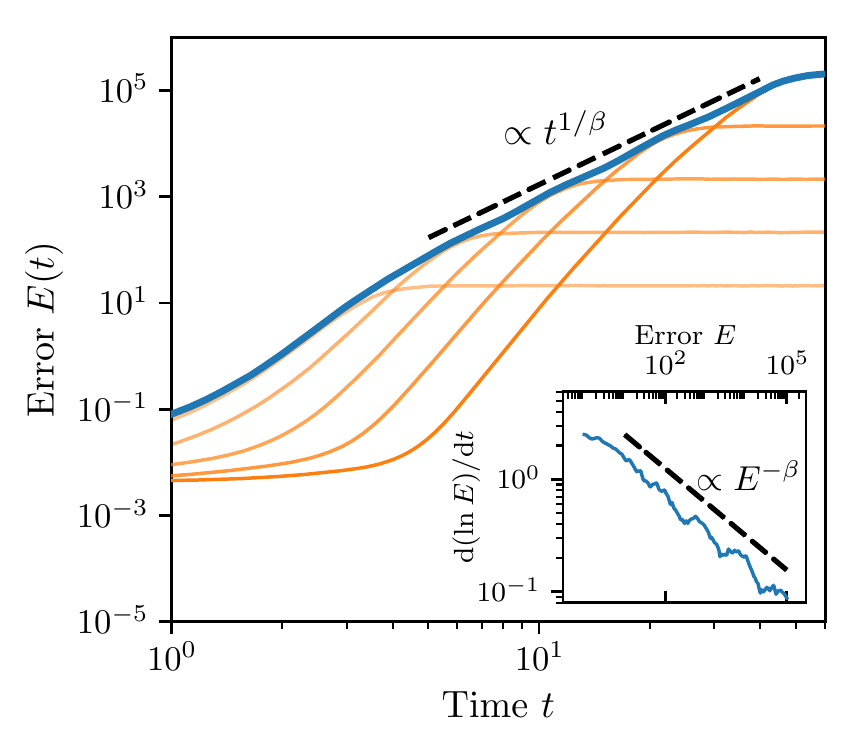}
	\caption{Power-law error growth of the model L96-H on a 
double-logarithmic plot}
	\label{fig:L96_H_5_power_law}
\end{figure}

We study a hierarchy of  $ L = 5 $ levels with the scale factors
$ \tau_i = 2^{i-5}$  and $ \alpha_i = 10^{5-i}$.
 The random initial perturbation has magnitude
$E_0=10^{-2}$ while the saturation value is at $E_{\infty,1}\approx
E_\infty\alpha_1 = 22\cdot 10^4$. 
In Figure  \ref{fig:L96_H_5_power_law} the error growth is shown on a
\textit{double logarithmic} plot. Both, the total error $E_{\rm tot}(t)$ 
(blue) and the error
growth of the levels $E_i(t)$ show power-law behavior, where the errors
measured in the sub-spaces of a certain level have additional features to be
discussed elsewhere. For the resulting power law error growth, the interplay
of level-$i$-Lyapunov exponents $\lambda \tau_i$ and of their 
saturation scales $E_{\infty,i} \approx E_\infty\alpha_i$ is crucial. 
The inset shows the numerically determined error
growth rates as a function of error magnitude.
The parameters $ \tau_i $ and $
\alpha_i $ are chosen such that the error growth rate decreases by a factor of
$ 1/2 $ every time the error becomes larger by a factor of $ 10 $. This
proportionality is shown by the bold dashed line with $ \drm(\ln E)/\drm t
\propto E^{-\ln(2)/\ln(10)} $. 

\section{Re-analysis of the Harlim et al. results}
\begin{figure}
\includegraphics[width=0.9\columnwidth]{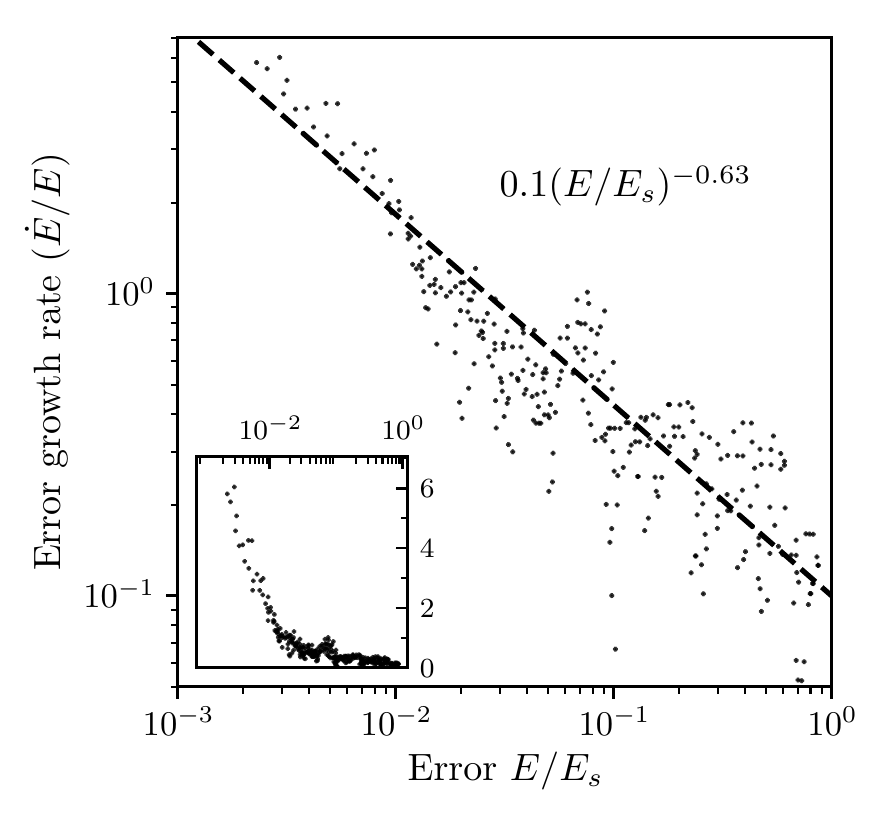}
\caption{\label{fig:Harlimdata}The dots are taken from Figure 1 of Harlim et
  al.\cite{Harlim2005} and denote error growth rate in units of 1/day 
as a function of error
  magnitude for a numerical weather model. The line is a power law fit with  
power $\beta=0.63$. The inset shows the scanned data in 
a representation as in the original
  publication and verifies that our recording of the plotted data is
  reasonable.}
\end{figure}

We have presented evidence that a power law divergence of error
growth rates can be realized by a dynamical system with properties which
resemble the observed coupling of time and length scales in the atmosphere. 
Here, we want to strengthen this concept by a study of a numerical weather
forecast system. Actually, since the authors of this article do neither have the
skills nor the resources to do a study on real weather forecasts, we use
published results to support our ideas. Harlim et al.\cite{Harlim2005} have
performed extensive numerical experiments with the Global Forecast System of the National Center
for Environmental Prediction (NCEP -GFS), focusing with their studies on
mid-latitudes wind prediction (vorticity). We recall here some details from
their original publication, for more see\cite{Harlim2005}.
 They applied perturbations to reference trajectories and measured
numerically the rate of divergence of such two trajectories as a function of
their Euclidian distance in phase space. The results are depicted in Figure 1 of
\cite{Harlim2005} as a scatter plot of error growth rate versus error
magnitude. We used the free software ``WebPlotDigitizer''
\cite{WebPlotDigitizer} to obtain the coordinates of an essential sub-set of
the dots in this diagram. These data, in the same representation as the
original figure (inset), and on a doubly logarithmic scale are shown in
Figure \ref{fig:Harlimdata}. Evidently, the error growth rate in this system can 
be well described by a power law with divergence for small errors
Equation (\ref{eq:2}), and the estimated power $\beta$ is about $0.63$. 
There is hence considerable evidence that a real weather forecast model 
does suffer not only from scale dependent error growth, but that this is 
indeed governed by a power law with a maximum prediction horizon of 15-16 days,
when we use Equation (\ref{eq:4}), 
insert $E_\infty=1$ (which is the saturation value in the
normalization of \cite{Harlim2005}), a time unit of days,
and $a=0.1$ and $\beta=0.63$ as obtained by our fit. 

\section{Conclusions}
Based on the idea of scale dependent error growth, which is motivated by  
meteorological evidence, we proposed the possibility of deterministic
dynamical systems with a strict prediction horizon, which is given if the
error growth rate diverges for small error magnitudes like a power law. 
We proposed a class of chaotic dynamical systems which exhibit such a
behavior and illustrated this by a model system. The system models the 
spatial and temporal hierarchies present in atmospheric dynamics. 
We then re-analyzed data produced in \cite{Harlim2005} in terms of 
a power law divergence of error growth rates and found thereby that
indeed in this weather forecast system, this power law divergence is present
and that the maximum forecast range is limited to 15 days. We find it 
plausible that the same holds for other weather forecast systems, as already
also stated for the European IFS\cite{Zhang2019}. 
The dynamical origin of this phenomenon lies in the linkage of spatial and 
temporal scales of this multi-scale phenomenon, where the smallest scales are 
the fastest.\\[.3cm]

\ack
We are grateful for fruitful discussions with M. Firmbach, C. Dubslaff, and 
B. L\"unsmann.

\end{document}